\documentclass{article}

\usepackage{microtype}
\usepackage{graphicx}
\usepackage{subcaption}
\usepackage{booktabs}
\usepackage{hyperref}

\usepackage[acceptedw]{icml2026}

\usepackage{amsmath}
\usepackage{amssymb}
\usepackage{mathtools}
\usepackage{amsthm}
\usepackage[capitalize,noabbrev]{cleveref}

\newcommand{\das}{\textsc{DAS}}
\providecommand{\clap}{}
\renewcommand{\clap}{CLAP}

\icmltitlerunning{Drift-Augmented Scoring for Noise-Robust Zero-Shot Audio Classification}

\begin{document}

\twocolumn[
  \icmltitle{Drift-Augmented Scoring: Text-Derived Noise Robustness \\
             for Zero-Shot Audio--Language Classification}

  \icmlsetsymbol{equal}{*}

  \begin{icmlauthorlist}
      \icmlauthor{Tu Vo}{ml2}
      \icmlauthor{Sheir Zaheer}{ml2}
      \icmlauthor{Chan Y. Park}{ml2}
    \end{icmlauthorlist}
    
    \icmlaffiliation{ml2}{KC Machine Learning Lab}
    
    \icmlcorrespondingauthor{Chan Y. Park}{chan.y.park@kc-ml2.com}

  \icmlkeywords{zero-shot audio classification, CLAP, noise robustness, test-time adaptation}

  \vskip 0.3in
]

\printAffiliationsAndNotice{}

\begin{figure*}[t]
  \centering
  \includegraphics[width=0.95\textwidth]{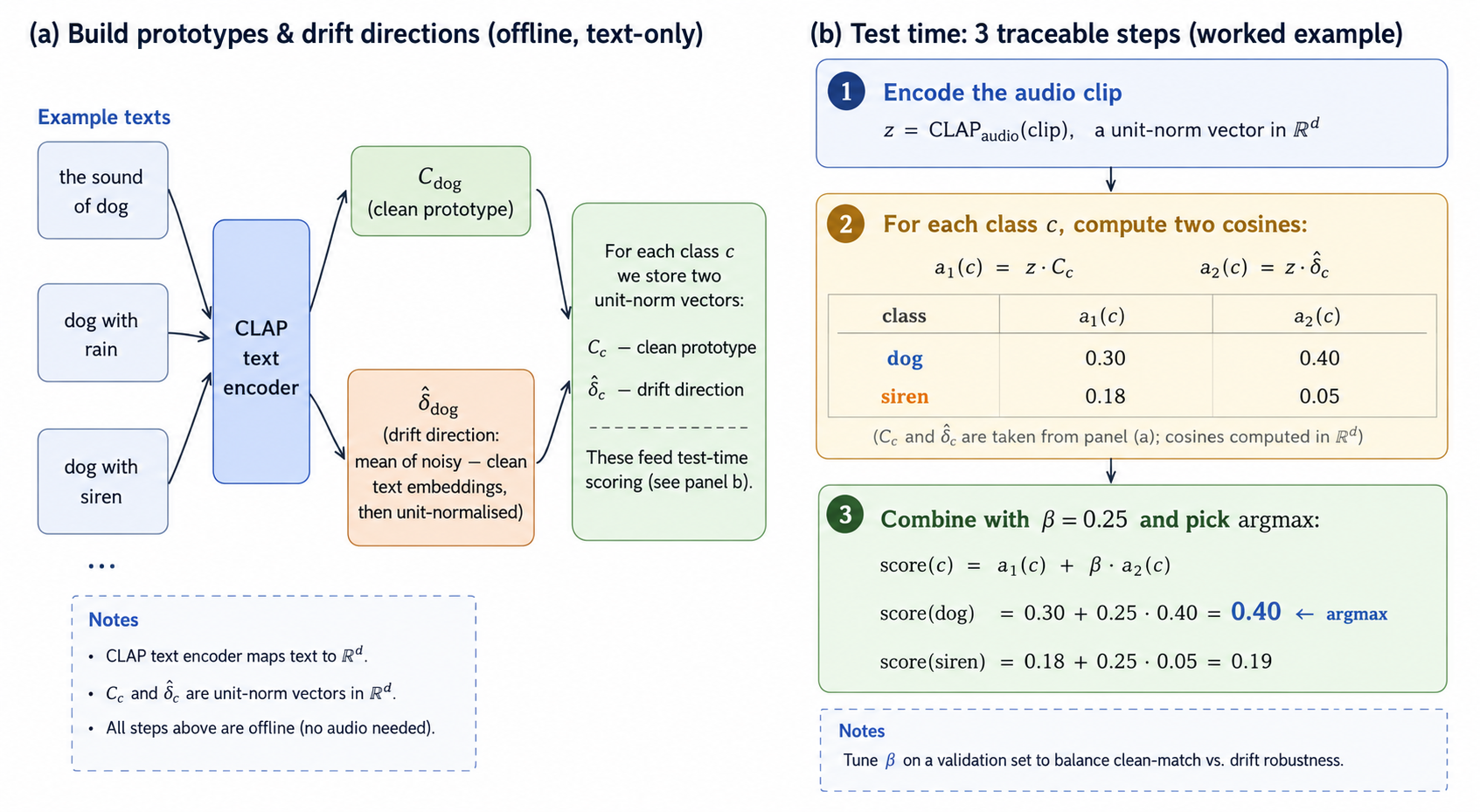}
  \caption{\textbf{Drift-Augmented Scoring (\das{}).}
    \textbf{(a)} Offline, text-only: for each class~$c$ the CLAP text
    encoder produces a clean prototype~$C_c$ from \texttt{"sound of
    }$c$\texttt{"} and a unit-norm drift direction~$\hat\delta_c$ that
    is the (normalised) mean of text drifts
    $\mathrm{CLAP}(\text{"}c\text{ with }p\text{"}) - C_c$ over a small
    set of noise phrases.
    \textbf{(b)} Test-time scoring in three steps:
    (1)~encode the audio clip into $z$;
    (2)~for each class $c$ compute two cosines---a clean cosine
    $a_1(c){=}z\!\cdot\!C_c$ and a drift cosine
    $a_2(c){=}z\!\cdot\!\hat\delta_c$;
    (3)~combine with weight~$\beta$ and pick $\arg\max_c$.
    Worked example shown is a noisy ``dog'' clip; full procedure
    in Algorithm~\ref{alg:das}.}
  \label{fig:das}
\end{figure*}

\begin{abstract}
Contrastive audio--language models such as \clap{} enable zero-shot
audio classification: a sound is labelled by matching its embedding
to text-prompt embeddings, with no labelled audio. This matching
breaks down under acoustic noise, where accuracy and mAP fall by
$12$--$30$ percentage points at $0$~dB SNR on standard benchmarks. We
propose \emph{Drift-Augmented Scoring} (\das{}), a small per-class
bonus added to the cosine score. The bonus rewards a class when the
noisy audio embedding drifts in the direction that the class's
noise-conditioned text prompts predict. It is derived from text
alone, computed once and cached, and adds a single inner product per
class at inference, with no gradients and no test-time batch. On a
LAION CLAP backbone, we compare \das{} against the four variants of
Acevedo et al.'s concurrent method on UrbanSound8K and the full
FSD50K eval, mixing each clip with urban acoustic scene noise across
a range of SNRs. \das{} improves the metric on every test
condition: by $+2.60$ to $+5.75$ accuracy points on UrbanSound8K and
$+1.50$ to $+1.74$~mAP on FSD50K.
\end{abstract}

\section{Introduction}

Pretrained contrastive audio--language models
(\clap{}~\cite{wu2023largescale,elizalde2023clap}) enable
\emph{zero-shot} audio classification. The user supplies class
names as text prompts, the text tower encodes them, and the
predicted class is the one whose prompt embedding is closest to the
audio embedding by cosine similarity. No labelled audio is needed.

This pipeline is fragile under noise.  Across standard sound-event
benchmarks (ESC-50, UrbanSound8K, FSD50K, SONYC-UST, DCASE~2016) and
real / synthetic noise corpora
(DEMAND~\cite{thiemann2013demand}, MUSAN~\cite{snyder2015musan}, TAU
Urban Acoustic Scenes 2019~\cite{mesaros2018tau}, additive Gaussian
noise with reverberation), zero-shot \clap{} accuracy/mAP drops by
$12$--$30$~pp at $0$~dB SNR relative to clean across every
configuration on a LAION CLAP backbone~\cite{wu2023largescale}.

Existing approaches to test-time adaptation in vision--language
models either modify the audio/image embedding directly via gradient
descent~\cite{shu2022tpt,wang2021tent}, operate
transductively~\cite{seth2025pat}, or iteratively
pseudolabel~\cite{hakim2025clipartt}. Subspace-based debiasing
methods~\cite{chuang2023debiasing,bi2025prism,zhu2024ssp} project
text embeddings off a global ``bias'' direction. The closest
concurrent work, Acevedo et al.~\cite{acevedo2025contextda}, either
subtracts a per-class bias from the cosine score or swaps the text
prototype for one retrieved from an in-domain audio pool
(Sec.~\ref{sec:related}). None of these derive a class-conditional
correction that depends on the noisy test clip without an audio
pool.

We propose \das{}: a one-line change to the zero-shot scoring rule
that adds a small text-derived bonus per class. Concretely:
\begin{equation}
  \mathrm{score}(z, c) \;=\; z \cdot C_c \;+\; \beta \cdot (z \cdot \hat\delta_c),
  \label{eq:das}
\end{equation}
where $\hat\delta_c$ is a unit-norm class-specific direction in
\clap{} space, derived from text by averaging the differences
\mbox{$\mathrm{CLAP}(\text{``}c\text{ with }p\text{"}) - C_c$} over a
small phrase set~$\mathcal{P}$ and normalising. The first term is
the standard zero-shot cosine; the second rewards classes whose
expected noise direction is consistent with how the noisy
audio~$z$ actually drifted (Figure~\ref{fig:das}).

\das{} has three properties:
\begin{itemize}\setlength\itemsep{1pt}
  \item \textbf{Class-conditional}: every class has its own drift
        direction~$\hat\delta_c$, derived from text-only drifts.
  \item \textbf{Audio-untouched}: $z$ is never modified, eliminating
        the confirmation-bias failure mode of iterative test-time
        adaptation~\cite{shu2022tpt,hakim2025clipartt}.
  \item \textbf{Single per-class score}: there is no $\max$ over
        candidate prototypes, so per-class ranking is preserved on
        multi-label tasks.
\end{itemize}


\section{Related Work}
\label{sec:related}

\paragraph{Noise-robust zero-shot CLAP.}
\clap{}~\cite{wu2023largescale,elizalde2023clap} and its successors
M2D-CLAP~\cite{niizumi2025m2dclap},
Audio~Flamingo~2~\cite{ghosh2025af2}, and
GLAP~\cite{dinkel2025glap} degrade under noise; \das{} modifies only
the scoring rule and so is a drop-in addition to any of them.  Prompt
engineering (ReCLAP~\cite{ghosh2025reclap},
TSPE~\cite{anand2025tspe}) and transductive prompt
re-weighting (PaT~\cite{seth2025pat}) improve clean accuracy but
either ignore noise or require the test batch.  Acevedo et
al.~\cite{acevedo2025contextda} is the closest concurrent
single-clip noise-robust work, with four variants in two families:
bias subtraction (ZS-Text, ZS-Audio), which removes a per-class but
\emph{clip-independent} profile from the cosine score, and text-guided
audio prototypes (TGAP, TGAP-Audio), which replace the class prototype
with embeddings retrieved from an in-domain audio pool; we compare
against all four in Sec.~\ref{sec:experiments}.  \das{} differs from
the first family by adding a correction that \emph{depends on the test
clip}, and from the second by needing no audio pool.

\paragraph{Test-time adaptation in multi-modal models.}
TPT~\cite{shu2022tpt}, TENT~\cite{wang2021tent}, and
CLIPArTT~\cite{hakim2025clipartt} adapt the model or embedding via
gradient steps or batched test data; MC-TTA~\cite{chen2024mctta} and
EMO-TTA~\cite{shi2025emotta} extend this style to \clap{};
CEA~\cite{liu2024cea} targets ASR foundation models; El Khoury et
al.~\cite{elkhoury2026entropy} re-weights prompts using prediction
entropy.  TPS~\cite{sui2025tps} shifts per-class prototypes at test
time for CLIP, single-sample.  Subspace-debiasing
methods~\cite{chuang2023debiasing,bi2025prism,zhu2024ssp} project a
\emph{global} bias direction out of class embeddings.  \das{} differs
from all of these by being closed-form, single-clip, gradient-free,
and producing a \emph{class-conditional} additive correction directly
from text.  Independently, CLAP-S~\cite{sun2025claps} uses the
acronym ``DAS'' for distributed acoustic sensing in fiber-optic
monitoring; the collision is purely lexical.

\section{Method}

\subsection{Setup}
\label{sec:setup}
A \clap{} model provides separate text and audio encoders that map
into a shared $d$-dimensional unit-norm embedding space.  For
zero-shot classification into a label set $\{1, \dots, C\}$, the
baseline encodes one prompt per class,
$\mathrm{prompt}(c) = $~\texttt{"the sound of }$c$\texttt{"}, to
obtain $C_c \in \mathbb{R}^d$, and predicts
$\hat{c} = \arg\max_c\ z\!\cdot\! C_c$ for an audio embedding $z$.
Acoustic noise corrupts $z$, displacing it from~$C_c$ and degrading
accuracy.  We apply \das{} on these drifted audio under different SNR level, 
and report gains in \emph{percentage points} (pp) of
accuracy or mean Average Precision (mAP) relative to baseline. 
All methods share the same encoders and normalization,
differing only in the scoring function.

\subsection{Text-derived drift direction}

Under noise, the audio embedding of a class-$c$ clip drifts along a
class-specific direction in \clap{} space. We hypothesise that the
\emph{text} representation of the same class, when conditioned on a
noise phrase, drifts in approximately the same direction. For each
class~$c$ and noise phrase~$p \in \mathcal{P}$, we measure the
text-side drift
\begin{equation}
  \mathbf{n}_{c,p,t} = \mathrm{CLAP}_{\text{text}}(T_{c,p}^{(t)})
                       - \mathrm{CLAP}_{\text{text}}(\text{"sound of }c\text{"})
\end{equation}
where $T_{c,p}^{(t)}$ instantiates one of $T{=}4$ composition
templates such as \texttt{"}$c$\texttt{ with }$p$\texttt{"}. With
$M{=}|\mathcal{P}|{=}52$ phrases and $T{=}4$ templates, this yields
$208$ drift vectors per class. We aggregate them into one direction
per class by simple averaging and normalisation:
\begin{equation}
  \hat\delta_c = \frac{\bar{\mathbf{n}}_c}{\|\bar{\mathbf{n}}_c\|},
  \qquad
  \bar{\mathbf{n}}_c = \frac{1}{TM} \sum_{t,p} \mathbf{n}_{c,p,t}.
\end{equation}

The phrase set~$\mathcal{P}$ is the union of three publicly
available noise descriptor lists (a synthetic-noise descriptor list,
the DEMAND environment list, and the MUSAN noise descriptors) plus
five generic phrases (\textit{``background noise'', ``ambient
sound'', ``noise in the recording'', ``audio interference'', ``low
quality audio''}). This is a generic noise vocabulary, not a
description of the test noise: the same set is used for every dataset
and every test-time corpus, with no per-corpus tuning. In particular,
it contains no TAU descriptors, so in the main experiments
(Sec.~\ref{sec:experiments}) the drift direction is built with no
knowledge of the test noise and must generalize to it.

\subsection{Drift-Augmented Scoring}

\das{} ranks classes by Eq.~\eqref{eq:das}. For single-label tasks
the prediction is $\arg\max_c \mathrm{score}(z, c)$; for multi-label
tasks the scores are used as per-class ranks for mean-AP computation.
The bonus term is not redundant with the cosine baseline: in general
$\hat\delta_c$ is not orthogonal to $C_c$, but the projection is
small. We sweep $\beta \in \{0.10, 0.25, 0.50\}$ and find that
$\beta{=}0.25$ gives the best mean gain on our panel (supplementary).
Algorithm~\ref{alg:das} summarises the full procedure.

\begin{algorithm}[t]
  \small
  \caption{Drift-Augmented Scoring (\das{}).}
  \label{alg:das}
  \begin{algorithmic}[1]
    \STATE \textbf{Inputs:} class labels $\{1,\dots,C\}$, prompt
      template $\mathrm{prompt}(c){=}$\texttt{"the sound of }$c$\texttt{"},
      noise-phrase set $\mathcal{P}$, $T$ composition templates,
      hyper-parameter $\beta$
    \STATE \textbf{Offline (once per class set, text-only):}
    \FOR{each class $c \in \{1,\dots,C\}$}
      \STATE $C_c \leftarrow \mathrm{normalize}\bigl(\mathrm{CLAP}_{\text{text}}(\mathrm{prompt}(c))\bigr)$
      \STATE collect text drifts
        $\mathbf{n}_{c,p,t} \!\leftarrow\! \mathrm{CLAP}_{\text{text}}(T_t(c,p)) - C_c$
        for all $p\in\mathcal{P}$, $t\in\{1,\dots,T\}$
      \STATE $\bar{\mathbf{n}}_c \leftarrow \mathrm{mean}_{p,t}\,\mathbf{n}_{c,p,t}$
      \STATE $\hat\delta_c \leftarrow \bar{\mathbf{n}}_c / \|\bar{\mathbf{n}}_c\|$
    \ENDFOR
    \STATE \textbf{Online (per test clip $x$):}
    \STATE $z \leftarrow \mathrm{normalize}\bigl(\mathrm{CLAP}_{\text{audio}}(x)\bigr)$
    \STATE $\mathrm{score}(c) \leftarrow z \cdot C_c \,+\, \beta \cdot (z \cdot \hat\delta_c)$ for $c=1,\dots,C$
    \STATE \textbf{return} $\arg\max_c \mathrm{score}(c)$ \quad
      (single-label) \quad or \quad
      $\bigl(\mathrm{score}(c)\bigr)_{c=1}^{C}$ \quad (multi-label rank)
  \end{algorithmic}
\end{algorithm}

\section{Experiments}
\begin{table*}[ht]
  \centering
  \footnotesize
  \renewcommand{\arraystretch}{1.05} 
  \setlength{\tabcolsep}{8pt}
  \caption{Per-SNR comparison on Scaper~\cite{salamon2017scaper}
    mixing with TAU Urban Acoustic Scenes 2019 backgrounds, on the
    LAION CLAP backbone (\texttt{larger\_clap}).  The top block is
    UrbanSound8K (single-label), scored by classification accuracy;
    the bottom block is the full FSD50K eval ($10{,}231$ clips,
    multi-label), scored by mean average precision (mAP).  All entries
    are percentages.  $\Delta$ is the gain of \das{} over the best of
    the four Acevedo et al.~\cite{acevedo2025contextda} variants at the
    same SNR.  \das{} wins all $10$ rows.}
  \label{tab:panel}
  \begin{tabular}{@{}rcccccccc@{}}
    \toprule
    SNR & Base & ZS-T-g & ZS-T-m & ZS-A-m & TGAP & TGAP-A & \textbf{\das{}} & $\Delta$ \\
    \midrule
    \multicolumn{9}{l}{\emph{UrbanSound8K (single-label, accuracy)}} \\
    $0$  & 54.80 & 40.30 & 40.80 & 39.85 & 53.40 & 52.45 & $\mathbf{59.15}$ & $\mathbf{+5.75}$ \\
    $6$  & 62.95 & 44.55 & 46.35 & 44.75 & 63.55 & 60.35 & $\mathbf{69.30}$ & $\mathbf{+5.75}$ \\
    $8$  & 65.95 & 46.85 & 47.30 & 45.75 & 65.60 & 63.40 & $\mathbf{71.15}$ & $\mathbf{+5.55}$ \\
    $10$ & 67.65 & 47.50 & 48.30 & 47.35 & 67.50 & 65.00 & $\mathbf{71.80}$ & $\mathbf{+4.30}$ \\
    $20$ & 72.90 & 51.85 & 51.55 & 50.40 & 74.40 & 72.20 & $\mathbf{77.00}$ & $\mathbf{+2.60}$ \\
    \midrule
    \multicolumn{9}{l}{\emph{FSD50K full $10{,}231$-clip eval (multi-label, mAP)}} \\
    $0$  & 33.31 & 33.31 & 33.31 & 33.31 & 30.23 & 30.23 & $\mathbf{35.03}$ & $\mathbf{+1.71}$ \\
    $6$  & 42.67 & 42.67 & 42.67 & 42.67 & 38.24 & 38.24 & $\mathbf{44.38}$ & $\mathbf{+1.71}$ \\
    $8$  & 44.83 & 44.83 & 44.83 & 44.83 & 40.25 & 40.25 & $\mathbf{46.57}$ & $\mathbf{+1.74}$ \\
    $10$ & 47.12 & 47.12 & 47.12 & 47.12 & 42.09 & 42.09 & $\mathbf{48.81}$ & $\mathbf{+1.69}$ \\
    $20$ & 54.88 & 54.88 & 54.88 & 54.88 & 48.76 & 48.76 & $\mathbf{56.38}$ & $\mathbf{+1.50}$ \\
    \bottomrule
  \end{tabular}
\end{table*}
\label{sec:experiments}

\textbf{Backbone.} We use the LAION CLAP
checkpoint~\cite{wu2023largescale} trained on music and speech
(HuggingFace \texttt{laion/larger\_clap\_music\_and\_speech}, denoted
\texttt{larger\_clap}). Sweeps over three other LAION CLAP backbones
(\texttt{clap-htsat-fused}, \texttt{clap-htsat-unfused},
\texttt{larger\_clap\_general}) are deferred to the supplementary.

\textbf{Datasets.} We use one single-label and one multi-label
benchmark, the largest of each kind in our sweep.  UrbanSound8K has
$10$ urban sound classes over $8{,}732$ evaluation clips; each clip
carries a single label, so we score it with classification accuracy.
FSD50K has $200$ classes over $10{,}231$ evaluation clips, with
several labels per clip, so we score it with mean average precision
(mAP).

\textbf{Noise corpus and mixing.} The background noise pool is TAU
Urban Acoustic Scenes 2019~\cite{mesaros2018tau}, restricted to the
six environment scenes used by Acevedo et
al.~\cite{acevedo2025contextda}: \emph{airport}, \emph{shopping
mall}, \emph{metro station}, \emph{park}, \emph{public square}, and
\emph{street traffic}. Each evaluation clip is mixed with a randomly
drawn TAU segment using the third-party
Scaper~\cite{salamon2017scaper} soundscape synthesiser. We evaluate
at $\{0, 6, 8, 10, 20\}$~dB, which covers Acevedo's reported
$\{6, 8, 10\}$~dB range and adds a heavy-noise ($0$~dB) and a mild
($20$~dB) endpoint. This gives a per-dataset $5$-SNR panel, or $10$
evaluation rows across the two datasets.

\textbf{\das{} configuration.} We fix $\beta{=}0.25$ for every result
here and sweep $\beta\in\{0.10, 0.50\}$ in the supplementary.  The
$52$-phrase set defined in the Method section is fixed across both
datasets and all SNRs; no per-dataset tuning.

\subsection{Main results}

We run all four Acevedo et al.~\cite{acevedo2025contextda} variants:
text- and audio-side bias subtraction (\textsc{ZS-Text},
\textsc{ZS-Audio}), and their text-guided audio prototypes
(\textsc{TGAP}, \textsc{TGAP-Audio}). The audio anchors use a
$200$-clip mean embedding from the matching noise corpus; TGAP uses
the top $N{=}10$ retrieved clean audio embeddings per class; the
bias coefficient is the paper-recommended $\tau{=}0.5$.
Table~\ref{tab:panel} reports per-SNR absolute scores. \das{} beats
every Acevedo variant on every row of UrbanSound8K ($5/5$, gaps
$+2.60$ to $+5.75$~acc) and FSD50K ($5/5$, gaps $+1.50$ to
$+1.74$~mAP). TGAP regresses below the baseline on every FSD50K
row, by $-3.08$ to $-6.13$~mAP. The text- and audio-bias variants
give \emph{exactly} the baseline mAP on FSD50K: their per-class
subtraction $z\cdot C_c - \tau k_c$ shifts every clip's score for
class~$c$ by the same constant~$k_c$, so the per-class ranking does
not change, and per-class AP does not change either. \das{} avoids
this degeneracy because $z\!\cdot\!\hat\delta_c$ depends on the clip.
\vspace{-1mm}
\subsection{Ablations}
\label{sec:ablation}

Each ablation below reports the mean gain ($\Delta$) averaged over the
ten rows of Table~\ref{tab:panel}: UrbanSound8K accuracy and FSD50K
mAP, each at the five SNRs.

\textbf{Mean vs.\ SVD top-$1$ drift.}  Per-class agreement between the
un-normalised mean drift and the top-$1$ singular direction of the
drift cloud is cosine $0.996$ (US8K) and $0.999$ (FSD50K), so a class's
$208$ drifts point in essentially one direction.  Swapping the mean for
the SVD direction moves the panel mean $\Delta$ by under $0.02$~pp
($+1.16$ vs $+1.18$, both positive on all $10$ rows), so we keep the
simpler mean.  

\textbf{Phrase set.}  The drift uses a $52$-phrase set (three published
noise-descriptor lists plus five generic phrases); rebuilding
$\hat\delta_c$ from only the five generic phrases drops the mean panel
gain from $+1.16$~pp ($10/10$ rows) to $+0.50$~pp ($5/10$).  The bonus
stays net-positive, so most of the signal is the bare concept of
``noise''.  The extra descriptors are not per-corpus tuning, since the
same set is used everywhere; they sharpen the averaged direction enough
to turn the gain positive on all ten rows.

\vspace{-2mm}
\section{Discussion and Limitations}

\paragraph{Compatibility with newer backbones.} \das{} touches only
the test-time scoring rule, not the encoders, so it transfers to
newer audio--language models (M2D-CLAP~\cite{niizumi2025m2dclap},
Audio~Flamingo~2~\cite{ghosh2025af2}, GLAP~\cite{dinkel2025glap})
without retraining, and composes with text-side scoring rules such
as entropy-based prompt re-weighting~\cite{elkhoury2026entropy}.
The phrase pool $\mathcal{P}$ is encoder-independent and the drift
vectors are cached, so the offline cost amortises across backbone
choices. Per-clip inference adds one inner product per class, orders
of magnitude below the audio-encoder forward pass.

\vspace{-2mm}
\paragraph{Does the text drift actually match the audio drift?}
\label{sec:modgap}
\das{} rests on the empirical bet that $\hat\delta_c$ tracks the
direction in which class-$c$ audio embeddings drift under noise.
For each row of Table~\ref{tab:panel} we compute
$\delta_{\text{audio},c} = \bar z_c^{(\text{noisy})}
- \bar z_c^{(\text{clean})}$ and its cosine with $\hat\delta_c$.
Pooled across UrbanSound8K ($10$ classes) and the full FSD50K eval
($200$ classes) at all five SNRs ($1{,}050$ class measurements), the
mean cosine is $+0.31$ and $99.0\%$ are positively aligned, with
FSD50K running at $98$--$100\%$ per SNR and UrbanSound8K at
$80$--$90\%$. The text encoder traces the same direction the audio
encoder takes under corruption, the empirical condition under which
Eq.~\eqref{eq:das} reduces to the right correction.
\vspace{-2mm}
\paragraph{Future work.} We fix $\beta{=}0.25$ across all panel
rows. Automatic per-class selection of $\beta$, analogous to
Acevedo et al.'s $\tau$ sweep~\cite{acevedo2025contextda}, is one
open direction; another is to refine $\hat\delta_c$ with an
unlabelled in-domain audio pool when one is available. A third is to
make the drift geometry backbone-aware: a short unlabelled
calibration set could align $\hat\delta_c$ with each encoder's
audio-side drift, since the strength of \das{} varies across
backbones in experiments. 

\vspace{-2mm}
\section{Conclusion}

\das{} adds a text-derived, class-conditional bonus to the cosine
score: $\mathrm{score}(z, c) = z\cdot C_c + \beta\,(z\cdot
\hat\delta_c)$. Following Acevedo et al.'s evaluation protocol
exactly (the \texttt{larger\_clap} backbone, TAU noise, Scaper
mixing, SNRs $\{0, 6, 8, 10, 20\}$~dB), \das{} beats every one of
their four variants on every one of the $10$ panel rows, raising UrbanSound8K accuracy from \(64.8\%\) to \(69.7\%\) and FSD50K mAP from \(44.6\) to \(46.2\) on average over SNRs . Within the panel, the text drift agrees in direction
with the audio drift on $99.0\%$ of $1{,}050$ class measurements at
mean cosine $+0.31$. Inference cost is one extra inner product per
class.

\bibliographystyle{icml2026}
\bibliography{dap}








\providecommand{\clap}{}
\renewcommand{\clap}{CLAP}

\icmltitlerunning{Drift-Augmented Scoring --- Supplementary Material}


\twocolumn[
  \icmltitle{Drift-Augmented Scoring:
             Text-Derived Noise Robustness for Zero-Shot
             Audio--Language Classification \\
             \large{\textsc{Supplementary Material}}}

  \icmlsetsymbol{equal}{*}

  \begin{icmlauthorlist}
    \icmlauthor{Anonymous Authors}{anon}
  \end{icmlauthorlist}

  \icmlaffiliation{anon}{Anonymous Institution}

  \icmlcorrespondingauthor{Anonymous}{anon@example.com}

  \icmlkeywords{zero-shot audio classification, CLAP, noise robustness, test-time adaptation}

  \vskip 0.3in
]

\printAffiliationsAndNotice{}

\noindent
This supplement collects analyses and tables omitted from the main
paper for space.  Section and table labels are prefixed with~\textsc{S}.
References to the main paper are by section number; the bib file is
shared.  Unless noted, $\Delta$ is the gain of \das{} ($\beta{=}0.25$)
over the zero-shot cosine baseline.

\section{Cross-backbone results}
\label{sec:backbones}

The main paper fixes the \texttt{larger\_clap} backbone.  To check that
\das{} is not tied to that choice, we run it on three further LAION
CLAP checkpoints~\cite{wu2023largescale}, including
\texttt{clap-htsat-fused}, the \texttt{630k-audioset-fusion}
checkpoint used by Acevedo et al.~\cite{acevedo2025contextda}.  Two
backbones (\texttt{larger\_clap}, \texttt{clap-htsat-fused}) are
evaluated on the main-paper noise setup (TAU backgrounds, Scaper
mixing); the other two were run under additive MUSAN, real-recorded,
and synthetic noise, so we report each backbone on the configuration we
have for it.  Table~\ref{tab:backbones} reports the per-dataset and
pooled gain for each backbone.

\begin{table*}[t]
  \centering
  \footnotesize
  \setlength{\tabcolsep}{5pt}
  \caption{\das{} ($\beta{=}0.25$) $\Delta$ over the cosine baseline on
    four LAION CLAP backbones, broken down by dataset.  Single-label
    datasets (ESC-50, UrbanSound8K, DCASE~2016) are scored with
    accuracy, multi-label datasets (SONYC-UST, FSD50K) with mAP.  The
    first two backbones use the main-paper noise setup (TAU backgrounds,
    Scaper mixing); the other two use additive MUSAN, real-recorded, and
    synthetic noise.  The last three columns pool over all of the
    backbone's evaluation rows.}
  \label{tab:backbones}
  \begin{tabular}{@{}lcccccccc@{}}
    \toprule
    & \multicolumn{5}{c}{Per-dataset mean $\Delta$}
    & \multicolumn{3}{c}{Pooled} \\
    \cmidrule(lr){2-6}\cmidrule(lr){7-9}
    Backbone & ESC-50 & US8K & DCASE & SONYC & FSD50K & Mean & Pos & Worst \\
    \midrule
    \texttt{larger\_clap}          & $+2.34$ & $+4.83$ & $+2.21$ & $+0.77$ & $+1.62$ & $\mathbf{+2.35}$ & $24/25$ & $-0.16$ \\
    \texttt{clap-htsat-fused}      & $+4.88$ & $+4.15$ & $+2.21$ & $+1.00$ & $+3.23$ & $\mathbf{+3.09}$ & $23/25$ & $-0.01$ \\
    \texttt{clap-htsat-unfused}    & $+0.49$ & $+1.37$ & $+2.62$ & $+0.29$ & $+1.28$ & $+1.21$          & $37/45$ & $-0.30$ \\
    \texttt{larger\_clap\_general} & $+0.99$ & $+1.63$ & $+0.51$ & $+0.76$ & $+0.76$ & $+0.93$          & $40/45$ & $-2.82$ \\
    \midrule
    \textbf{Total}                 & & & & & & $\mathbf{+1.66}$ & $\mathbf{124/140}$ & $-2.82$ \\
    \bottomrule
  \end{tabular}
\end{table*}

\das{} is net-positive on all four backbones: $124$ of $140$ rows,
mean $\Delta{=}{+}1.66$.  On \texttt{clap-htsat-fused}, Acevedo et
al.'s exact checkpoint, the gain reaches $+3.09$ over baseline, so the
method carries over to their backbone and not just ours.  The
per-dataset columns show the gain is largest on the single-label event
and scene sets, especially UrbanSound8K and ESC-50 on the two fused
backbones, and smallest on the $8$-class SONYC-UST, the same pattern as
Table~\ref{tab:methods}.  The two backbones with smaller pooled gains
were evaluated under additive noise that includes the synthetic-AWGN
corpus, the hardest case for the drift mechanism
(Sec.~\ref{sec:modgap}); that difference in noise setup, more than the
backbone itself, accounts for much of the spread.  This variation
across backbones motivates the backbone-aware calibration we flag as
future work in the main paper.

\section{Comparison with the baselines across the full grid}
\label{sec:methods}

The main paper compares \das{} with the four Acevedo et
al.~\cite{acevedo2025contextda} variants on the panel, where \das{}
wins all ten rows.  Table~\ref{tab:methods} widens that comparison to
the full $327$-row sweep (every backbone, dataset, corpus, mixing, and
SNR we ran), reporting each method's gain over the cosine baseline.

\begin{table*}[t]
  \centering
  \footnotesize
  \setlength{\tabcolsep}{4pt}
  \caption{Each method's $\Delta$ over the cosine baseline across the
    full $327$-row sweep, split into single-label accuracy and the two
    multi-label mAP datasets.  The methods behave oppositely by task
    type: bias subtraction (ZS-Text, ZS-Audio) leaves mAP \emph{exactly}
    unchanged and hurts accuracy; TGAP helps accuracy but loses on mAP,
    collapsing on SONYC-UST (negative on all $66$ rows, worst
    $-17.2$); \das{} is the only method positive across all three, with
    by far the smallest worst case.}
  \label{tab:methods}
  \begin{tabular}{@{}lccccccccc@{}}
    \toprule
    & \multicolumn{3}{c}{Accuracy ($198$ rows)}
    & \multicolumn{3}{c}{mAP: FSD50K ($63$)}
    & \multicolumn{3}{c}{mAP: SONYC-UST ($66$)} \\
    \cmidrule(lr){2-4}\cmidrule(lr){5-7}\cmidrule(lr){8-10}
    Method & Mean $\Delta$ & Pos & Worst & Mean $\Delta$ & Pos & Worst & Mean $\Delta$ & Pos & Worst \\
    \midrule
    ZS-Text (generic) & $-8.86$ & $26/198$  & $-23.72$ & $0.00$  & $0/63$  & $0.00$  & $0.00$  & $0/66$  & $0.00$ \\
    ZS-Text (matched) & $-7.69$ & $16/198$  & $-21.47$ & $0.00$  & $0/63$  & $0.00$  & $0.00$  & $0/66$  & $0.00$ \\
    ZS-Audio          & $-1.72$ & $89/198$  & $-22.50$ & $0.00$  & $0/63$  & $0.00$  & $0.00$  & $0/66$  & $0.00$ \\
    TGAP              & $+3.98$ & $152/198$ & $-13.16$ & $-1.59$ & $11/63$ & $-4.17$ & $-7.96$ & $0/66$  & $-17.22$ \\
    TGAP-Audio        & $+3.41$ & $142/198$ & $-10.51$ & $-1.59$ & $11/63$ & $-4.17$ & $-7.96$ & $0/66$  & $-17.22$ \\
    \textbf{\das{}}   & $\mathbf{+2.41}$ & $\mathbf{177/198}$ & $\mathbf{-2.82}$ & $\mathbf{+1.72}$ & $\mathbf{63/63}$ & $\mathbf{+0.32}$ & $\mathbf{+0.95}$ & $\mathbf{58/66}$ & $\mathbf{-0.27}$ \\
    \bottomrule
  \end{tabular}
\end{table*}

Two patterns explain the panel result.  The bias-subtraction variants
(ZS-Text, ZS-Audio) leave multi-label mAP \emph{exactly} unchanged on
both datasets (their per-class constant cancels under ranking), and on
single-label accuracy they mostly hurt (ZS-Text $-7.7$ mean, positive
on $16/198$ rows), because subtracting a fixed profile moves every
class the wrong way as often as the right way.  TGAP is the mirror
image: strong on single-label accuracy (mean $+3.98$, $152/198$), where
retrieving in-domain audio prototypes sharpens fine-grained sets such as
ESC-50, but it loses on multi-label mAP.  The loss is mild on FSD50K
($-1.59$) and severe on the $8$-class SONYC-UST, where it is negative on
\emph{every} one of the $66$ rows (mean $-7.96$, worst $-17.2$):
swapping each class prototype for a retrieved audio mean destroys the
per-class score calibration that average precision depends on.  \das{}
is the only method positive on both task types ($+2.41$ accuracy, and
$+1.72/{+}0.95$ mAP on FSD50K\,/\,SONYC-UST), and the only one whose
worst case anywhere stays within $-3$.  A per-row oracle that always
picked the best Acevedo variant would edge \das{} on raw mean, yet
\das{} still beats that unrealisable oracle on $184/327$ rows, and
every TGAP variant needs an in-domain audio pool that \das{} does
without.  \das{} gives up TGAP's occasional large wins in exchange for
not regressing badly on any condition.

\section{Robustness to noise type, mixing, and SNR}
\label{sec:robustness}

\das{} adds one text-derived direction that is never tuned to the
test-time noise, yet the gain holds across every noise condition we
tried.  Table~\ref{tab:robust} breaks it down three ways: by noise
corpus, by mixing method, and by SNR.

\begin{table*}[t]
  \centering
  \footnotesize
  \setlength{\tabcolsep}{6pt}
  \caption{\das{} ($\beta{=}0.25$) $\Delta$ over the cosine baseline
    across the full $327$-row grid, broken down by noise corpus, by
    mixing method, and by SNR.  \das{} is positive in every group.}
  \label{tab:robust}
  \begin{tabular}{@{}lccccc@{}}
    \toprule
    & Cells & Mean $\Delta$ & Pos & Worst & Best \\
    \midrule
    \multicolumn{6}{l}{\emph{By noise corpus}} \\
    TAU            & $80$ & $\mathbf{+2.55}$ & $73/80$ & $-0.45$ & $+6.40$ \\
    DEMAND         & $65$ & $+2.44$ & $61/65$ & $-1.54$ & $+5.70$ \\
    MUSAN          & $95$ & $+1.62$ & $84/95$ & $-2.82$ & $+4.70$ \\
    real-recorded  & $42$ & $+1.54$ & $39/42$ & $-1.03$ & $+4.89$ \\
    synthetic      & $45$ & $+1.52$ & $41/45$ & $-0.30$ & $+3.75$ \\
    \midrule
    \multicolumn{6}{l}{\emph{By mixing method}} \\
    Scaper         & $150$ & $+2.46$ & $143/150$ & $-1.54$ & $+6.40$ \\
    additive       & $162$ & $+1.46$ & $141/162$ & $-2.82$ & $+5.35$ \\
    additive-TAU   & $15$  & $\mathbf{+2.88}$ & $14/15$ & $-0.11$ & $+5.50$ \\
    \midrule
    \multicolumn{6}{l}{\emph{By SNR}} \\
    $0$~dB  & $89$ & $+1.77$ & $77/89$ & $-1.54$ & $+6.40$ \\
    $6$~dB  & $30$ & $\mathbf{+2.58}$ & $30/30$ & $+0.33$ & $+6.35$ \\
    $8$~dB  & $30$ & $+2.46$ & $29/30$ & $-0.26$ & $+5.50$ \\
    $10$~dB & $89$ & $+1.87$ & $80/89$ & $-1.54$ & $+5.60$ \\
    $20$~dB & $89$ & $+1.96$ & $82/89$ & $-2.82$ & $+5.40$ \\
    \bottomrule
  \end{tabular}
\end{table*}

By corpus, the gain is largest on structured real-world noise (TAU and
DEMAND, both above $+2.4$) and smallest on synthetic AWGN ($+1.52$),
the same ordering as the drift alignment in Sec.~\ref{sec:modgap}: the
more the test noise resembles a describable acoustic environment, the
better the text-derived drift tracks it.  By mixing, the result
survives both simple additive mixing and the Scaper soundscape
pipeline.  By SNR, the mean stays between $+1.8$ and $+2.6$ from $0$ to
$20$~dB, so the gain does not depend on heavy noise: \das{} helps about
as much at a mild $20$~dB as at $0$~dB, and can stay on as a default
scoring rule instead of being switched in only when noise is known to
be severe.

\section{Modality-gap correlation across backbones and corpora}
\label{sec:modgap}

The premise of \das{}, that the text-side $\hat\delta_c$ approximates
the direction in which a class-$c$ audio embedding moves under noise,
is empirical.  The main paper measures it on the panel
(Sec.~5: $1{,}050$ class measurements, mean cosine ${+}0.31$, $99.0\%$
positive).  Here we widen the test to three backbones
(\texttt{larger\_clap}, \texttt{clap-htsat-unfused},
\texttt{larger\_clap\_general}), two single-label datasets (ESC-50,
UrbanSound8K), and three additive noise corpora (MUSAN, real-recorded,
synthetic) at $0$~dB.  For each row we compute the audio-side per-class
drift
\begin{equation*}
  \delta_{\text{audio},c} \;=\; \bar z_c^{(\text{noisy},\,0\,\text{dB})}
                                   \;-\; \bar z_c^{(\text{clean})}
\end{equation*}
from class-conditional means and measure its cosine with $\hat\delta_c$.
Pooled over $540$ class measurements, the mean cosine is $+0.287$ with
$96.9\%$ of classes positively aligned.  MUSAN and real-recorded noise
align best (mean $+0.345$, $99\%$ positive), while synthetic AWGN is
weakest (mean $+0.172$, $92\%$ positive).  Across both the panel and
this broader sweep, the text encoder traces, on average, the same
direction the audio encoder takes under corruption: the empirical basis
for the main paper's Eq.\,(1).

\section{Per-dataset behaviour}
\label{sec:per-dataset}

\begin{table*}[t]
  \centering
  \footnotesize
  \setlength{\tabcolsep}{6pt}
  \caption{\das{} ($\beta{=}0.25$) $\Delta$ over the cosine baseline per
    dataset, pooled over the full grid (all backbones, noise corpora,
    mixings, and SNRs in which the dataset appears).  Single-label
    datasets are scored with accuracy, multi-label with mean average
    precision.}
  \label{tab:per-dataset}
  \begin{tabular}{@{}llccccc@{}}
    \toprule
    Dataset      & Type                  & Metric & Cells & Mean $\Delta$ & Pos    & Worst \\
    \midrule
    UrbanSound8K & events, 10 cls        & acc    & $66$  & $\mathbf{+3.41}$ & $66/66$ & $+0.10$ \\
    DCASE~2016   & scenes, 15 cls        & acc    & $66$  & $+2.00$          & $56/66$ & $-2.82$ \\
    ESC-50       & events, 50 cls        & acc    & $66$  & $+1.84$          & $55/66$ & $-0.60$ \\
    FSD50K       & multi-label, 200 cls  & mAP    & $63$  & $+1.72$          & $63/63$ & $+0.32$ \\
    SONYC-UST    & multi-label, 8 cls    & mAP    & $66$  & $+0.95$          & $58/66$ & $-0.27$ \\
    \bottomrule
  \end{tabular}
\end{table*}

\noindent
\das{} is positive on every dataset (Table~\ref{tab:per-dataset}),
single-label and multi-label alike.  UrbanSound8K is the strongest
case ($+3.41$, positive on all $66$ rows) and FSD50K the clearest
multi-label one ($+1.72$ on all $63$).  The per-class additive bonus
preserves per-class ranking, which is why \das{} scales to long-tailed
multi-label sets where alternative scoring rules (prompt tuning,
score-bias subtraction) regress; see the main paper's Table~1.

\section{Choice of the drift weight $\beta$}
\label{sec:beta}

\das{} has a single hyperparameter, the weight $\beta$ on the drift
bonus in Eq.\,(1) of the main paper.  We fix $\beta{=}0.25$ everywhere
and never tune it per dataset.  Table~\ref{tab:beta} sweeps it over the
ten panel rows (UrbanSound8K accuracy and FSD50K mAP at the five
SNRs).  At $\beta{=}0.10$ the bonus is positive on every row but too
small to use the drift fully; at $\beta{=}0.25$ the mean gain peaks and
stays positive on all ten rows; at $\beta{=}0.50$ the bonus begins to
overwhelm the cosine on fine-grained classes, the mean gain drops, and
half the rows turn negative.  The gain is stable in a band around
$0.25$ and degrades only once the bonus is large enough to compete with
the base score.

\begin{table}[t]
  \centering
  \footnotesize
  \setlength{\tabcolsep}{6pt}
  \caption{Drift weight $\beta$ swept over the ten panel rows
    (UrbanSound8K accuracy and FSD50K mAP at five SNRs).  $\beta{=}0.25$
    gives the best mean gain and is positive on every row.}
  \label{tab:beta}
  \begin{tabular}{@{}ccccc@{}}
    \toprule
    $\beta$ & Mean $\Delta$ & Pos & Worst & Best \\
    \midrule
    $0.10$          & $+0.61$          & $10/10$          & $+0.20$          & $+0.91$ \\
    $\mathbf{0.25}$ & $\mathbf{+1.16}$ & $\mathbf{10/10}$ & $\mathbf{+0.35}$ & $\mathbf{+1.74}$ \\
    $0.50$          & $+0.15$          & $5/10$           & $-1.41$          & $+1.60$ \\
    \bottomrule
  \end{tabular}
\end{table}

\section{Safety: no confirmation bias}
\label{sec:safety}

A common failure mode of test-time prompt or feature adaptation is
confirmation bias: an iterative update trained on the model's own
(possibly wrong) prediction reinforces it, especially when the
unsupervised signal is weak under noise.  \das{} avoids this entirely.
Its scoring rule has no gradient loop and never mutates the audio
embedding~$z$; the drift directions $\hat\delta_c$ are computed once,
offline, from text, with no information from the test clip.  Under heavy
noise the worst case is that the $z\!\cdot\!\hat\delta_c$ bonus becomes
uninformative and the rule sits close to the baseline cosine.  Because
nothing is updated from the model's own output, a regression on one clip
cannot compound across clips; across the full grid the largest
single-row regression anywhere is $-2.82$ and the largest gain
$+6.40$ (Table~\ref{tab:robust}).


\end{document}